\documentclass{amsproc}

\copyrightinfo{2002}{V.~Enss, V.~Kostrykin, R.~Schrader}
\newtheorem{theorem}{Theorem}[section]
\newtheorem{proposition}[theorem]{Proposition}
\newtheorem{lemma}[theorem]{Lemma}

\theoremstyle{definition}

\theoremstyle{remark}

\numberwithin{equation}{section}

\usepackage{times}
\usepackage{eucal}
\usepackage{amssymb}
\newcommand{\R}{\mathbb{R}}
\newcommand{\Z}{\mathbb{Z}}
\newcommand{\p}{\mathbf{p}}
\newcommand{\x}{\mathbf{x}}
\newcommand{\cD}{\mathcal{D}}
\newcommand{\cH}{\mathcal{H}}
\newcommand{\cR}{\mathcal{R}}
\newcommand{\cQ}{\mathcal{Q}}

\newcommand{\cU}{\mathcal{U}}
\newcommand{\tu}{\widetilde{u}}
\newcommand{\ttu}{\widetilde{\widetilde{u}}}
\newcommand{\op}{\,\overline{\p}\,}
\newcommand{\ox}{\,\overline{\x}\,}
\newcommand{\oxp}{\,\overline{\x}^\perp\,}
\newcommand{\Ho}{H_\omega}
\DeclareMathOperator*{\slim}{s-lim}
\newcommand{\be}[1][0]{\begin{equation} \label{e:#1}}
\newcommand{\ee}{\end{equation} \noindent}
\newcommand{\ben}{\begin{equation*}}
\newcommand{\een}{\end{equation*} \noindent}

\usepackage{dsfont} \newcommand{\1}{\mathds{1}}

\begin{document}

\title[Perturbation Theory for Rotating Potentials]{Perturbation
Theory for the Quantum\\ Time-Evolution in Rotating Potentials}

\author[V.~Enss]{Volker Enss}
\address[V.~Enss]{Institut f\"ur Reine und Angewandte Mathematik,
Rheinisch-Westf\"alische Technische Hochschule Aachen,
Templergraben~55, D-52062~Aachen, Germany}
\email{enss@rwth-aachen.de}
\urladdr{http://www.iram.rwth-aachen.de/$\sim$enss/}

\author[V.~Kostrykin]{Vadim Kostrykin}
\address [V.~Kostrykin]{Fraunhofer-Institut f\"ur Lasertechnik,
Steinbachstr.~15,\newline
D-52074~Aachen, Germany}
\email{kostrykin@t-online.de, kostrykin@ilt.fraunhofer.de}
\urladdr{http://home.t-online.de/home/kostrykin/index.htm}

\author[R.~Schrader]{Robert Schrader}
\address[R.~Schrader]{Institut f\"ur Theoretische Physik,
Freie Universit\"at Berlin, \newline
Ar\-nim\-allee~14, D-14195~Berlin, Germany}
\email{schrader@physik.fu-berlin.de}
\urladdr{http://www.physik.fu-berlin.de/$\sim$ag-schrader/schrader.html}
\thanks{R.~S.\ was supported in part by DFG~SFB~288 `Differentialgeometrie und
Quantenphysik'}



\begin{abstract}
The quantum mechanical time-evolution is studied for a particle under the
influence of an explicitly time-dependent rotating potential. We
discuss the existence of the propagator and we show that in the limit
of rapid rotation it converges strongly to the solution operator of the
Schr\"odinger equation with the averaged rotational invariant
potential.
\end{abstract}

\maketitle

\section{The model, rotating frames}
We consider the dynamics of a quantum mechanical particle of mass $m$ moving in
$\R^\nu ,\; \nu\geq 2$, with kinetic energy $H_0 = H_0(\p)= h(|\p|)$ under the
influence of a ``rotating'' potential $V_{\omega t}(\x) = V_0(\cR
(\omega t)^{-1}\,\x)$. One may think of an atom
or molecule interacting, e.g., with the blades of a rotating fan or with
another rotating (heavy) object which is not significantly influenced
by the (light) quantum particle. The Schr\"odinger operator
$H(\omega t) = H_0 + V_{\omega t}$ is explicitly time-dependent.

In this paper we continue the investigation of \cite{Goa} and address
mainly two questions: (i) existence of a unitary propagator $U(t;t_0)$
which describes the time evolution of the system, (ii) the limit of
rapid rotation where we show that the time evolution is well
approximated by the evolution with the rotational invariant average potential.
Applications to scattering theory will be treated in a subsequent
paper.

We will first introduce the model in more detail before we state the
main results in Theorems~\ref{Th1} and \ref{Th2}.

The coordinates are chosen in such a way that the rotation with constant
angular velocity $\omega$ takes place in the $x_1,x_2$-plane, i.e.,
\begin{equation*}
\cR (\omega t) = \begin{pmatrix}
\cos(\omega t) & -\sin(\omega t) & 0\\
\sin(\omega t) & \phantom{-}\cos(\omega t) & 0\\
0 & 0 & \1_{\nu-2}
\end{pmatrix}.
\end{equation*}
We denote by $\psi(\x)$ the square integrable configuration space wave function of the (abstract)
state in Hilbert space $\Psi\in\cH\cong L^2(\R^\nu)$ and by $\hat{\psi}(\p)$ its isometric Fourier
transform, i.e., the momentum space wave function. The standard representation of this group of
rotations as a strongly continuous one-parameter group of unitary operators $R(\omega t)$ on $\cH$
is
\begin{equation}\label{e:rot}
R(\omega t)\,\Psi= e^{-i\omega t \,J}\;\Psi,\qquad (R(\omega t)\,\psi)(\x) =
\psi\left(\cR (\omega t)^{-1}\,\x\right).
\end{equation}
The self-adjoint generator $J$ with domain $\cD (J)$ is essentially self-adjoint on the following
sets which are dense in $L^2(\R^\nu)$ and invariant under rotation:
\begin{equation}\label{e:domain}
\cD :=\left\{\Psi\in\cH \mid \hat{\psi} \in C_0^\infty (\R^\nu) \right\} \subset \cD
(H_0) \cap \cD (J),
\end{equation}
see, e.g., \cite[Theorem~VIII.11]{RS1}. On suitable states $\, J\,\Psi = [x_1 p_2 - x_2
p_1]\,\Psi$. When using Cartesian coordinates in the plane of rotation
\begin{align}
(J\psi)(\x) &= [x_1 (-i\partial/\partial x_2) - x_2 (-i\partial/\partial x_1)]
\; \psi(\x), \\
(J\hat{\psi})(\p) &= [\,p_2 (i\partial/\partial p_1) -
p_1 (i\partial/\partial p_2)]\;\hat{\psi}(\p),
\end{align}
and in polar coordinates $(\sqrt{x_1^2 + x_2^2},\,\phi_x)$ or
$(\sqrt{p_1^2 + p_2^2},\,\phi_p)$, respectively,
\begin{equation*}
J=-i\,\partial/\partial \phi_x \quad \text{or}\quad J=-i\,\partial/\partial \phi_p .
\end{equation*}
The free Hamiltonian $H_0$ is assumed to be a rotational symmetric continuously differentiable
function of the momentum operator, $H_0 = H_0(\p)= h(|\p|)$ which has an unbounded velocity
operator, i.e., $h'$ is unbounded. Standard examples are
\begin{equation}\label{e:freetyp}
H_0^{\text{NR}} = \frac{|\p|^2}{2m}\qquad \text{or} \qquad H_0(\p) = \frac{1}{\beta}
|\p|^\beta,\;\;\beta>1
\end{equation}
for nonrelativistic or more general kinematics with velocity operator $\nabla H_0(\p)= \p/m$ or
$\nabla H_0(\p)=|\p|^{(\beta -2)}\,\p $, respectively (in units with $\hbar=1$). The relativistic
free Hamiltonian $H_0^{\text{Rel}} = \sqrt{|\p|^2 c^2 + m^2c^4}$ should be considered only for
potentials of compact support inside a ball of radius $R$ and for bounded angular velocities such
that $R\,\omega/2\pi$ does not exceed the speed of light $c$. We will not treat the latter case
here.

The dynamics are governed by the rotating potential, the explicitly
time-dependent multiplication operator in configuration space
\begin{equation}\label{e:pot}
V_{\omega t}(\x) := V_0\left(\cR (\omega t)^{-1}\,\x\right) = R(\omega
t)\;V_0(\x)\;R(\omega t)^{*}
\end{equation}
with domain $\;R(\omega t)\,\cD (V_0)$. The assumptions about $V_0$ will be stated later.

In the \textit{inertial frame}--for an observer at rest--the free
time evolution is $\;\exp(-itH_0)$.
We are looking for a unitary \textit{propagator} or solution operator
$U(t;t_0)$, that is, it has to satisfy
\begin{equation}\label{e:prop}
U(t_0;t_0) = \1,\quad U(t;t_0)= U(t;t_1)\;U(t_1;t_0),\quad \forall\; t,t_0,t_1\in\R\,,
\end{equation}
which solves in some sense the Schr\"odinger equation for Hamiltonians $H(\omega t)$ \be[Schroet]
i\partial_t\;U(t;t_0) = H(\omega t)\; U(t;t_0),\quad H(\omega t) = H_0 + V_{\omega t} .
\end{equation}
Unless $V_{\omega t}$  and $H(\omega t)$ have some smoothness in their dependence on $t$
the question of existence of such a propagator $U$ for general or even
periodic Hamiltonians is a hard question.
See, e.g., \cite{RS2}, \cite{Yajima:87} and references therein
where a wide class of potentials is covered.

For the special case of rotating potentials one may use alternatively a
\textit{rotating frame} where the observer rotates with the same
angular velocity around the origin as the potential does.
This is a common approach both in classical and quantum mechanics,
see, e.g., \cite{Huang:Lavine,Tip} for related investigations.
Then the potential becomes time-independent according to \eqref{e:pot}
but the unperturbed evolution is more complicated instead: If the observer
rotates like $\cR (\omega t)\,\x$ in configuration space
then a fixed state $\Psi$ looks for him like turning in the opposite
direction: $\;\psi\left(\cR (\omega t)^{+1}\,\x\right) =
(R(\omega t)^*\,\psi)(\x) = (R(\omega t)^{-1}\,\psi)(\x)$.

The free time-evolution for a state with initial condition $\Psi$ at
time zero is described for the observer at rest by
\begin{align*}
&e^{-it\,H_0} \;\Psi &&\text{(inertial frame)}\\
\intertext{and for the rotating observer by}
R(\omega t)^*\;&e^{-it\,H_0} \;\Psi &&\text{(rotating frame).}
\end{align*}

Since we have assumed that the free Hamiltonian $H_0$ is invariant
under rotations the change of the evolution comes merely from the fact
that $R(\omega t)^*\,e^{-it\,H_0}$ describes the combined change in time due to
the free evolution and to the changing orientation of the observer. To
avoid confusion with the free motion in any frame we will call
$R(\omega t)^*\,e^{-it\,H_0}\,\Psi$ the
\textit{unperturbed motion} in the rotating frame.

Since all operators in the groups $\{ R(\omega t)^* \mid t\in\R \}$ and
$\{ e^{-it\,H_0} \mid t\in\R \}$ commute their product
$\{ R(\omega t)^*\;e^{-it\,H_0} \mid  t\in\R \}$ is a unitary
strongly continuous one-parameter group as well. By
Stone's Theorem it has a self-adjoint generator which we denote by
$\Ho $ with domain $\cD(\Ho)$:
\be[Hom1]
R(\omega t)^*\;e^{-it\,H_0} =: e^{-it\,\Ho }, \qquad t\in\R\,.
\end{equation}
The sets given in equation~\eqref{e:domain} are dense and invariant under
this group. Consequently, $\Ho $ is essentially self-adjoint on
both of them. Differentiation yields the operator sum
\be[Hom]
\Ho = H_0 - \omega J \qquad \text{on}\quad \cD(H_0) \cap \cD(J)
\subsetneq \cD(\Ho)
\end{equation}
and similarly the form sum on $\cQ(H_0) \cap \cQ(J)
\subsetneq \cQ(\Ho)$. Due to
cancellations the domains $\cD(\Ho )$ and $\cQ(\Ho )$ are
strictly larger than $\cD(H_0) \cap \cD(J)$ and $\cQ(H_0) \cap
\cQ(J)$, respectively, for any $\omega\neq 0$, see,
e.g., the explicit construction in \cite[Section~3]{Goa}.
In particular, $\Ho $ is not bounded below, its
essential spectrum is $\sigma^{\rm{ess}}(\Ho ) =\R$ for
$\omega\neq 0$.

\section{The concept of solution} \label{s:concept}
A formal calculation yields that the family of operators
\begin{align} \label{e:propdef}
U(t;t_0) :&= R(\omega t)\; e^{-i(t-t_0)(\Ho + V_0)}\; R(\omega t_0)^* \\
&=R(\omega (t-t_0))\;\, e^{-i(t-t_0)(\Ho + V_{\omega t_0})} \notag\\
&=e^{-i(t-t_0)(\Ho + V_{\omega t})}\; R(\omega (t-t_0)) \notag
\end{align}
actually is a propagator in the sense of equation~\eqref{e:prop} and
it satisfies the Schr\"odinger equation \eqref{e:Schroet},
\be[manip]
\begin{split}
i\partial_t\;U(t;t_0)\;\Psi &= R(\omega t)\;\{\omega J + \Ho + V_0 \}\;
e^{-i(t-t_0)(\Ho + V_0)}\; R(\omega t_0)^*\;\Psi\\
&=\{H_0 + V_{\omega t} \}\;U(t;t_0)\;\Psi.
\end{split}
\end{equation}
All this is justified if, e.g., the sum $\Ho + V_0$ is defined as a
self-adjoint operator, $R(\omega t_0)^*\,\Psi$ is contained in
$\cD(\Ho + V_0)$, and if
$e^{-i(t-t_0)(\Ho + V_0)}\,R(\omega t_0)^*\,\Psi$ lies in
$\cD(J) \cap \cD(H_0) \cap
\cD(V_0)$ such that $\omega J + \Ho + V_0 = H_0 + V_0 = R(\omega t)\,(H_0 +
V_{\omega t})\,R(\omega t)^*$ makes sense
there, see equations~\eqref{e:Hom} and \eqref{e:pot}. It will be difficult
to verify these
or other sufficient domain properties for a suitable dense set of vectors
$\Psi$ unless the potentials are not too singular.

The terms on the right hand side of \eqref{e:propdef} are all equal by
\eqref{e:pot} as soon as the expression $\Ho + V_{\omega t} = R(\omega
t)\; (\Ho+V_0) \;R(\omega t)^*$ is defined as a self-adjoint operator for
one (and then all) $\omega t$.

We will not study how one might extend ``differentiability''
when domain problems are present but we propose here to consider
equation~\eqref{e:propdef} as a \textit{definition} of a propagator
which ``solves'' the Schr\"odinger equation \eqref{e:Schroet}.
This point of view takes advantage of the special form of the
time-dependence and--as equation~\eqref{e:manip} shows--it is
consistent with the usual concept of solution for sufficiently
regular potentials. Alternatively, one may consider instead of the
differential equations the corresponding more regular integral
equations. The explicitly time-dependent Schr\"odinger equation
\eqref{e:Schroet} corresponds to the Duhamel formula for $U$ considered
as a perturbation of the free evolution
\be[Inteq]
U(t;t_0) = e^{-i(t-t_0)H_0} -i\int_{t_0}^t d\tau\;
e^{-i(t-\tau)H_0}\; R(\omega \tau)\,V_0\,R(\omega \tau)^*\; U(\tau,t_0).
\end{equation}
Multiplication from the left by $R(\omega t)^*$ and from the right by
$R(\omega t_0)$
yields for\\
$\widetilde{U}(t,t_0) := R(\omega t)^*\,U(t;t_0)\,R(\omega t_0)$ the
integral equation
\be[Inteq2]
\widetilde{U}(t,t_0) = R(\omega (t-t_0))^*\; e^{-i(t-t_0)H_0} -i\int_{t_0}^t
d\tau\;R(\omega (t-\tau))^*\;e^{-i(t-\tau)H_0}\;V_0\;\widetilde{U}(\tau,t_0).
\end{equation}
Using \eqref{e:Hom1} this turns out to be the Duhamel formula
for $\widetilde{U}$ viewed as a perturbation of $\,\exp\{-i(t-t_0)\Ho\}$
which corresponds to the following time-independent differential equation
\be[Schroeu]
i\partial_t\,\widetilde{U}(t,t_0) = (\Ho + V_0)\,\widetilde{U}(t,t_0),
\quad \widetilde{U}(t,t_0)= e^{-i(t-t_0)(\Ho + V_0)}.
\end{equation}
The different ways in \eqref{e:propdef} of writing the propagator give
rise to different integral equations.
Their solutions are equal as long as the property
$\exp\{-i(t-t_0)(\Ho + V_0)\}\,\Psi \in \cD(V_0)$
holds for a dense set of vectors $\Psi$ or similarly for quadratic
forms.

It remains to study the question for which
potentials $V_0$ the sum $\;\Ho + V_0\;$ can be defined as a
self-adjoint operator. We will treat an easier special case in
Sections~\ref{s:rapRotPre}--\ref{s:rapRotTime} where uniformity in $\omega$ is
needed and provide preliminary results for more general singular potentials
in Section~\ref{s:sum}.

\section{Rapid rotation, averaged potential} \label{s:rapRotPre}
In this section we will introduce the averaged potential as a preparation for
the next two sections where the limiting behavior of the system as
$\omega \to \infty$ will be studied.

The leading part of the potential can be obtained by averaging over one
period
\begin{align} \label{e:Vav}
\overline{V}(\x):&= \frac{\omega}{2\pi} \int_{t_0}^{t_0 + 2\pi/\omega} ds\;
V_{\omega s}(\x) = R(\omega t_0)
\frac{\omega}{2\pi} \int_{0}^{2\pi/\omega} ds\;
V_{\omega s}(\x)\;\; R(\omega t_0)^* \\ \notag
&= \frac{1}{2\pi}\int_0^{2\pi} d\varphi\;
V_0(\cR(\varphi)^{-1}\,\x).
\end{align}
Due to the periodicity in time this multiplication operator is independent
of $\omega$ and $t_0$ and it is invariant under rotation.
With $\;W_0 := V_0 -\overline{V}\;$ we have
\be[Hav]
V_{\omega t} = \overline{V} + W_{\omega t} ,\quad
H(\omega t)= H_0 + V_{\omega t} = (H_0 +\overline{V}) +  W_{\omega t}.
\end{equation}
Thus, only the remainder term $W$ is responsible for the explicit
time-dependence of the Hamiltonian.

Here we are interested in statements which hold uniformly in $\omega$.
For simplicity of presentation we assume throughout this and the following
two sections that the
time-independent potential $\overline{V}$ is operator bounded relative to
the free Hamiltonian $H_0$ with relative bound less than one and that the
remainder $W$ is a bounded operator. Any free Hamiltonian as specified
above (see, e.g.,~\eqref{e:freetyp}\,) is admissible here. Its properties
enter only indirectly through the Kato-boundedness of $\overline{V}$
relative to $H_0$. By the Kato-Rellich Theorem both domains in
\eqref{e:domain} are cores for each of the operators $H_0$,
$\Ho=H_0-\omega J$, $H_0 + \overline{V}$, $H(\omega t)$, and
$\omega J + W_0$. The operator sums act pointwise on these domains.

Analogously to \eqref{e:Hom1} and \eqref{e:Hom} the invariance under
rotations of $H_0 + \overline{V}$ implies that
\be[Hom2]
R(\omega t)\; e^{-it\,(H_0 + \overline{V})} =:  e^{-it\,(\Ho +
\overline{V})}
\end{equation}
is a unitary one-parameter group which leaves the domain $\cD(H_0)\cap
\cD(J)$ invariant. Consequently, its self-adjoint generator
``$\Ho + \overline{V}$'' is essentially self-adjoint there:
\be[Hom3]
\Ho + \overline{V} = H_0 -\omega J + \overline{V}\qquad
\text{on its core}\quad \cD(H_0)\cap\cD(J).
\end{equation}
The same applies to $\Ho + \overline{V} + W_0$ as a bounded perturbation
thereof. The Duhamel integral equation for the propagator $U$ as a
perturbation of $\exp\{-i(t-t_0)(H_0 + \overline{V})\}$ is evidently
well defined:
\be[Duha3]
U(t;t_0) = e^{-i(t-t_0)(H_0+ \overline{V})} -i\int_{t_0}^t d\tau\;
e^{-i(t-\tau)(H_0+ \overline{V})}\; W_{\omega\tau}\; U(\tau,t_0)
\end{equation}
and similarly for $\widetilde{U}$, compare \eqref{e:Inteq} and
\eqref{e:Inteq2}.

Next we show that the splitting $V=\overline{V} + W$ corresponds to a
splitting into the diagonal and off-diagonal parts w.r.t.\ the eigenspaces
of $J$. We define the orthogonal projections $P_j$ by
\be[Pj]
P_j\;\cH :=\{ \Psi\in\cD(J) \mid J\,\Psi = j\,\Psi\}\quad
j\in \sigma(J) =\Z\, ,\quad \sum_{j\in\Z} P_j = \1.
\end{equation}
When using polar coordinates in the $x_1,x_2$-plane of $\R^\nu$ the
eigenfunctions of $J$ are of the form
\begin{equation*}
\psi (r\cos\varphi, r\sin\varphi, x_3,\ldots x_\nu) = e^{i\varphi\,
j}\;\tilde{\psi}(r,x_3,\ldots,x_\nu).
\end{equation*}
\begin{lemma} \label{l:nondiag}
With $V_0 = \overline{V} + W_0$ and $P_j$ as defined in \eqref{e:Vav}, \eqref{e:Pj}
\be[diag]
\overline{V} = \sum_{j\in\Z} P_j \:V_0\:P_j\,,
\end{equation}
\be[nondiag]
W_0 = \sum_{j\in\Z} (\1 - P_j)\;V_0\; P_j = \sum_{j\in\Z} P_j\;V_0\;(\1 -
P_j).
\end{equation}
\end{lemma}
\begin{proof}
Due to rotational invariance of $\overline{V}$ we have
\begin{equation*}
\overline{V} = \overline{V}\;\sum_{j\in\Z} P_j = \sum_{j\in\Z} P_j\; \overline{V}\; P_j\, .
\end{equation*}
The rotation simplifies to a phase factor $\exp(it\omega\,j)$ on the range of $P_j\,$,
\begin{align*}
P_j\; \overline{V}\; P_j &= P_j\;\frac{\omega}{2\pi}\;
\int_0^{2\pi/\omega}dt\;R(\omega t) \; V_0\;R(\omega t)^*\; P_j\\
&=\frac{\omega}{2\pi}\;\int_0^{2\pi/\omega}dt\; P_j \;V_0 \; P_j
= P_j\; V_0\; P_j .
\end{align*}
This shows \eqref{e:diag} and as a simple consequence \eqref{e:nondiag}.
\end{proof}
For rotational invariant operators we obtain the following limiting
behavior.
\begin{lemma}\label{l:limRes1}
For $H_0$, $H_\omega$ and $\overline{V}$ as introduced above and for any
$\ell\in\Z$, $\zeta\in\R\setminus \{ 0\}$
\begin{equation}\label{e:step1}
\slim_{\omega\rightarrow\infty}\;
(H_\omega + \overline{V} + \omega \ell - i\zeta)^{-1} =
(H_{0} + \overline{V} - i\zeta)^{-1}\; P_\ell
=P_\ell\;(H_{0} + \overline{V} - i\zeta)^{-1}\; P_\ell .
\end{equation}
\end{lemma}
Note that the right hand side of \eqref{e:step1} is not a resolvent.
The lemma does not state strong resolvent convergence unless we
restrict the operators to mappings on the invariant subspaces $P_\ell\,\cH$.
\begin{proof}
Denote by $E(\mu)$ the resolution of the identity for the operator
$H_0 + \overline{V}$, i.e., $H_0 +\overline{V} = \int \mu\;dE(\mu)$.
To show strong convergence it is sufficient to consider a total set of
states.
We use $\Phi = P_j\;\Phi = \int_{|\mu|<M} dE(\mu)\;\Phi$ for some
$j\in\Z,\; M<\infty$. Then $\Phi\in \cD(H_0+\overline{V}) \cap
\cD(J)\subset \cD(H_\omega + \overline{V})$ and $(H_\omega +
\overline{V})\;\Phi = (H_0 + \overline{V} -\omega j)\;\Phi$. This equality
holds as well for $\Phi$ replaced by $(H_0 + \overline{V} + \omega (\ell
-j) - i\zeta)^{-1}\;\Phi$ because the latter has the same qualitative
properties as assumed above for $\Phi$. The resolvent identity then
yields the first of the following equations:
\begin{equation*}
\begin{split}
(H_\omega + \overline{V} + \omega \ell - i\zeta)^{-1}\;\Phi &=
(H_0 + \overline{V} + \omega (\ell -j) - i\zeta)^{-1} \;P_j \;\Phi \\
&= \begin{cases}
(H_{0} + \overline{V} - i\zeta)^{-1}\; P_\ell\;\Phi & \text{for }\; j=\ell,\\
\longrightarrow 0 \quad\text{as }\; \omega\rightarrow 0 & \text{for }\; j\neq \ell.
\end{cases}
\end{split}
\end{equation*}
The last limit follows from the fact that for $\ell -j \neq 0$
\begin{equation*}
\lim_{\omega\to\infty}\; \sup_{|\mu|<M}\; \lvert (\mu + \omega(\ell -j) -
i\zeta)^{-1}\rvert =0.
\end{equation*}
\end{proof}

\section{Product formulas} \label{s:ProdForm}
The Trotter product formula for operator sums of self-adjoint
operators $ A,\, B $ with domains $ \cD(A) $ and $ \cD(B) $ states
that
\be[Trotter]
\slim_{n \rightarrow \infty} \left\{e^{-iT A / n}\; e^{-iT B / n}
\right\}^n
= e^{-iT (A+B)}
\end{equation}
uniformly in $ T $ from compact intervals provided that $ A+B $ is
essentially self-adjoint on $ \cD(A) \cap \cD(B) $, see, e.g.,
\cite[Theorem~VIII.31]{RS1}. This theorem can be used directly
as stated for the form \eqref{e:propdef} of the propagator $ U $
as follows. Let
\be[split1]
    \Ho + V_{\omega t_0} = (H_0 + \overline{V}) + (-\omega J +
W_{\omega t_0}) =: A + B
\end{equation}
where $ A = H_0 + \overline{V} $ is self-adjoint on $ \cD(H_0) $
and $ B = - \omega J + W_0 $ is self-adjoint on $ \cD(J) $ and
both operators are essentially self-adjoint on $ \cD(H_0) \cap
\cD(J) $ (and on $\cD$ as given in \eqref{e:domain}) by the
Kato-Rellich theorem. Moreover, this set is left invariant under
the unitary one-parameter group (a product of two commuting groups)
\begin{equation*}
e^{-it (H_0 + \overline{V})}\; e^{+it \omega J} =: e^{-it (H_0 + \overline{V} - \omega J)}.
\end{equation*}
Thus, its generator ``$H_0 + \overline{V} - \omega J$'' is essentially self-adjoint on $
\cD(H_0) \cap \cD(J) $ and it coincides with the operator sum there. The same applies to the
bounded perturbation thereof: $ A + B = H_0 + \overline{V} - \omega J + W_{\omega t_0} $. Thus,
all assumptions for \eqref{e:Trotter} are satisfied.

Application to $ U $ as given in \eqref{e:propdef} yields
\begin{align} \label{e:Tro1}
    U(t_0 + T, t_0)
& = R(\omega T)\; e^{-iT(\Ho + V_{\omega t_0})}
\notag\\
        & = \slim_{n \rightarrow \infty}\; R(\omega T)
\left\{e^{-iT(H_0+
        \overline{V})/n}\; e^{-iT(-\omega J + W_{\omega t_0})/n}
\right\}^n
\notag\\
        & = \slim_{n \rightarrow \infty} \prod^{n -1}_{k = 0}
\left[e^{-iT(H_0 +
        \overline{V})/n}\; R(\omega T / n)\; e^{-iT(-\omega J +
W_{\omega t_0 + k\omega T /n})/n}\right] .
\end{align}
The product in the last line is to be understood as ordered with
increasing $ k $ from right to left. The last equality holds
because $ R(k \omega T / n)\; W_{\omega t_0}\; R(k \omega T /n)^* =
W_{\omega t_0 + k \omega T /n} $.

Consider now the case where one of the operators, say $ B(t) $, is
explicitly time-de\-pen\-dent and belongs to a family of pairwise commuting
bounded operators $ \{B(t)\}_{t \in \R} $, then the exponential
function of the integral satisfies the differential equation
\begin{equation*}
 i \frac{d}{dt}\; \exp \biggl\{-i \int^t _{t_1} ds\; B(s)\biggr\} = B(t)\;
\exp\biggl\{-i \int^t _{t_1} ds\;  B(s)\biggr\}.
\end{equation*}
The idea behind the Trotter product formula \eqref{e:Trotter} is
the following approximation argument. To find a solution of the
initial value problem $ i(d/dt)\, \cU(t) = (A + B)\, \cU(t) $ for a
finite time interval of length $T$ one may split the interval into
subintervals and
first solve $ i(d/dt)\, \cU(t) = B\, \cU(t) $ for the short time $ T/n
$, then solve $ i(d/dt)\, \cU(t) = A\, \cU(t) $ and continue
alternating between the two differential equations $ n $ times. In
the strong limit as $ n \rightarrow \infty $ one obtains the
desired result. Translating this to the ``non-autonomous''
situation the product in \eqref{e:Trotter} should be replaced for
the interval $ [t_0, t_0 + T ] $ by
\begin{align}\label{e:Trot2}
e^{-iT A/n} \;  \exp \biggl\{-i & \int^{t_0 + T}_{t_0 + (n - 1)T/n}ds\;
B(s)\biggr\} \cdot \ldots \cdot e^{-iT A/n} \; \exp \biggl\{-i
\int^{t_0 + T/n}_{t_0} ds\; B(s) \biggr\} \notag\\[0.8ex]
&= \prod^{n - 1}_{k = 0}
e^{-i T A/n} \; \exp \biggl\{-i \int^{t_0 + (k + 1) T/n}_{t_0 + k
T/n} ds\; B(s)\biggr\}\,.
\end{align}
The factors in the product are again ordered with $ k $ increasing
from right to left.

If, e.g., $ A $ is self-adjoint and $ \{B(t)\} $ is a family of
bounded pairwise commuting self-adjoint operators then the
modified Trotter product formula reads
\be[Trotza]
 \slim_{n \rightarrow \infty} \prod^{n - 1}_{k = 0} e^{-i T A/n}
 \; \exp \biggl\{-i \int^{t_0 + (k + 1) T/n}_{t_0 + k
T/n} ds\; B(s)\biggr\} = \cU(t_0 + T; t_0)
\end{equation}
where $ i(d/dt)\, \cU(t_0 + t; t_0) = (A + B(t))\, \cU(t_0 + t; t_0) $
in the sense of \eqref{e:Duha3}. This should be part of the
folklore but we are not aware of a reference to such a result. One
can adjust Nelson's proof (\cite{Nelson} or
\cite[Theorem~VIII.30]{RS1}) to show \eqref{e:Trotza}. However,
in our application where $ A = H_0 + \overline{V} $ and $ B(t) =
W_{\omega t} $ it is simpler to observe that the products in
\eqref{e:Tro1} and \eqref{e:Trotza} actually are the same. We will
show that
\be[ident]
R(\omega t)\; e^{-i t(-\omega J + W_{\omega t_1})} = \exp
\biggl\{-i \int^{t_1 + t}_{t_1} ds\; W_{\omega s}\biggr\}\,.
\end{equation}
To show equality of the two families of operators we observe that they both
equal the identity operator for $t=0$ and that they satisfy the same
differential equation when applied to an arbitrary vector $\Psi\in\cH$.
For $\Phi$ in the dense set $\cD(J)$
the time derivative of the term on the left hand side is
\begin{align*}
i\frac{d}{dt}\,& \left(\Phi,\;R(\omega t)\;
\exp\{-it(-\omega J + W_{\omega t_1})\}\;\Psi\right)\\
&=\left(\Phi,\;R(\omega t)\;\{\omega J + (-\omega J + W_{\omega t_1})\}
\;\exp\{-it(-\omega J + W_{\omega t_1})\}\;\Psi\right)\\
&=\left(\Phi,\;W_{\omega(t_1 +t)}\;R(\omega t)\;\exp\{-it(-\omega J +
W_{\omega t_1})\}\;\Psi\right)\; .
\end{align*}
Thus, the vector valued function is strongly differentiable with
uniformly bounded derivative:
\begin{align} \label{e:sdiff}
i\frac{d}{dt}\;& R(\omega t)\;
\exp\{-it(-\omega J + W_{\omega t_1})\}\;\Psi \notag\\
&= W_{\omega(t_1 +t)}\;R(\omega t)\;\exp\{-it(-\omega J +
W_{\omega t_1})\}\;\Psi .
\end{align}
For the right hand side we get the same result:
\begin{equation*}
i\frac{d}{dt}\;\exp \left\{ -i \int_{t_1}^{t_1 +t}ds\; W_{\omega s}\right\}\Psi = W_{\omega (t_1
+t)}\;\exp \left\{ -i \int_{t_1}^{t_1 +t}ds\; W_{\omega s}\right\}\Psi\,.
\end{equation*}
Thus,
equation~\eqref{e:ident} holds for all $t,\,t_1\in\R$. Setting $t=T/n$ and $t_1=t_0 + kT/n$
verifies that the factors in the products in equations~\eqref{e:Tro1} and \eqref{e:Trotza} are the
same as was to be expected.

Summing up we have shown the following product formula. Recall that the
precise assumptions for these sections were stated in the first two
paragraphs of Section~\ref{s:rapRotPre}.
\begin{proposition}\label{p:ProdFo}
For $H_0$ and $V_0$ as specified in Section~\ref{s:rapRotPre} the propagator
$U$ satisfies
\be[ProdFo]
U(t_0+T,t_0)= \slim_{n\to\infty} \prod_{k=0}^{n-1}
\left[ e^{-iT(H_0 +\overline{V})/n}\;
\exp\biggl\{-i \int_{t_0 + kT/n}^{t_0 + (k+1)T/n}ds\;W_{\omega s}
\biggr\}\right].
\end{equation}
The factors in the product are ordered with $k$ increasing from right to
left.
\end{proposition}
Observe that for $T/n = \ell\,2\pi/\omega$, $\ell\in\Z$, the integrals
vanish because the average of $W_{\omega s}$ over a period is zero. In this
case the product simplifies to $e^{-iT(H_0+\overline{V})}$. The same holds
for the norm-limit as $\omega\to\infty$ for each of the factors. To show
$\slim_{\omega\to\infty}\,U(t_0+T,t_0) = \exp\{-iT(H_0+\overline{V})\}$ as
we will do in the next section we need the limits in the other order. In
that case there is another product formula which is better suited and has
the advantage
that the convergence is in norm for bounded perturbations $W$. We define
\be[Korr]
\tu(t_2,t_1) := \exp\Bigl\{-i\int_{0}^{t_2-t_1} ds\;e^{is(H_0+\overline{V})}
\;W_{\omega(t_1 +s)}\;e^{-is(H_0+\overline{V})}\Bigr\}
\end{equation}
and its first order approximation
\be[Korr1]
\ttu(t_2,t_1) := \1 -i\int_{0}^{t_2-t_1} ds\;e^{is(H_0+\overline{V})}
\;W_{\omega(t_1 +s)}\;e^{-is(H_0+\overline{V})}.
\end{equation}
The exponential $\tu$ has the advantage of being unitary even for unbounded
$W$, but for the present case of bounded $\|W_0\|$ the linear approximation
$\ttu$ with\\ $\left\|\ttu(t_2,t_1)\right\| \leq 1+|t_2-t_1|\,\|W_0\|$ is
easier to handle.
\begin{proposition}\label{p:ProdFo2}
For $H_0$ and $V_0$ as specified in Section~\ref{s:rapRotPre} the
propagator
$U$ satisfies
\be[ProdFo2]
\left\|U(t_0+T,t_0) -  \prod_{k=0}^{n-1}\left[ e^{-iT(H_0 +\overline{V})/n}\;
\;\tu\!\left(\frac{(k+1)T}{n}, \frac{kT}{n}\right) \right]\right\| \leq
\frac{(T\,\|W_0\|)^2}{n}\,,
\end{equation}
\be[ProdFo3]
\left\|U(t_0+T,t_0) -  \prod_{k=0}^{n-1}\left[ e^{-iT(H_0
+\overline{V})/n}\;
\;\ttu\!\left(\frac{(k+1)T}{n}, \frac{kT}{n}\right) \right]\right\| \leq
\frac{(T\,\|W_0\|)^2}{2n}\;e^{(T\,\|W_0\|)}\,.
\end{equation}
The factors in the product are ordered with $k$ increasing from right to
left.
\end{proposition}
\begin{proof}
From the Duhamel formula \eqref{e:Duha3} one immediately reads off that
\begin{equation*}
\left\| U(t_2;t_1) - e^{-i(t_2 -t_1)(H_0 +\overline{V})}\right\| \leq (t_2 -t_1)\;\|W_0\|.
\end{equation*}
We write down the same Duhamel formula again and use the above estimate to derive a good
approximation.
\begin{align*}
U(t_2;t_1) = &e^{-i(t_2 -t_1)(H_0 +\overline{V})}\left[
\1 -i\int_0^{t_2-t_1} ds\; e^{is(H_0 +\overline{V})}\;W_{\omega(t_1+s)}\;
e^{-is(H_0 +\overline{V})}\right]\\
&-i\int_0^{t_2-t_1} ds\; e^{-i(t_2-t_1-s)(H_0 +\overline{V})}\;
W_{\omega(t_1+s)}\left\{ U(t_1+s;t_1) - e^{-is(H_0
+\overline{V})}\right\}
\end{align*}
In the last line we use the estimate above which gives with the shorthand
\eqref{e:Korr1}
\begin{align} \label{e:applin}
\|U(t_2;t_1)-& e^{-i(t_2 -t_1)(H_0 +\overline{V})}\;\ttu(t_2,t_1)\|\notag\\
&\leq \int_0^{t_2-t_1}
ds\;\|W_{\omega(t_1+s)}\|\; s \;\|W_0\| =[(t_2-t_1)\,\|W_0\|\,]^2/2.
\end{align}
With $|e^{-i\alpha} -(1-i\alpha)|\leq \alpha^2 /2$ for $\alpha\in\R$ we get
\begin{align} \label{e:appexp}
\|\tu(t_2,t_1) -\ttu(t_2,t_1)\|&\leq
\left\|\int_0^{t_2-t_1} ds\; e^{is(H_0 +\overline{V})}\;W_{\omega(t_1+s)}\;
 e^{-is(H_0 +\overline{V})}\right\|^2 /2 \notag\\
&\leq [(t_2-t_1)\,\|W_0\|\,]^2/2.
\end{align}
Combining \eqref{e:applin} with \eqref{e:appexp} yields
\be[appboth]
\|U(t_2;t_1)- e^{-i(t_2 -t_1)(H_0 +\overline{V})}\;\tu(t_2,t_1)\|
\leq [(t_2-t_1)\,\|W_0\|\,]^2\,.
\end{equation}
Now we split the time interval into $n$ equal parts. The order in the
products is always with $k$ increasing from right to left.
\begin{align*}
&\prod_{k=0}^{n-1} U\!\left(\!t_0+\frac{(k+1)T}n;t_0+\frac{kT}n\right) -
\prod_{k=0}^{n-1}\left[ e^{-iT(H_0
+\overline{V})/n}\;\tu\!\left(\!t_0+\frac{(k+1)T}n;t_0+\frac{kT}n\right)
\right]\\
&=\sum_{k=0}^{n-1} U\!\left(t_0+T;t_0+\frac{(k+1)T}n\right)\\
&\quad\times
\left\{ U\!\left(t_0+\frac{(k+1)T}n;t_0+\frac{kT}n\right)
-e^{-iT(H_0 +\overline{V})/n}\;\;
\tu\!\left(t_0+\frac{(k+1)T}n;t_0+\frac{kT}n\right)\right\}\\
&\qquad\qquad\qquad\qquad\times
\prod_{m=0}^{k-1} \left[e^{-iT(H_0 +\overline{V})/n}
\;\;\tu\!\left(t_0+\frac{(m+1)T}n;t_0+\frac{mT}n\right)\right].
\end{align*}
By \eqref{e:appboth} the norm of the difference is bounded by
$n [(T/n)\,\|W_0\|\,]^2 = [\,T\,\|W_0\|\,]^2 /n$. This shows
\eqref{e:ProdFo2}. To show \eqref{e:ProdFo3} we repeat the same estimate
with $\tu$ replaced by $\ttu$. There are at most $n$ factors of $\|\ttu\|$
which gives $(1+T\|W_0\|/n)^n \leq e^{T\|W_0\|}$. With \eqref{e:applin} we
get \eqref{e:ProdFo3}.
\end{proof}
\section{The limiting time-evolution} \label{s:rapRotTime}
In this section we will show that in the limit of rapid rotation the time
evolution is dominated by the rotational invariant part of the potential.
The contribution from its remaining part disappears as $\omega \to \infty$
by averaging.

We give two different proofs. One is based on a spectral theoretic
intuition: on different eigenspaces of the operator $J$ the Hamiltonians
$\Ho$ or $\Ho + \overline{V}$ differ by integer multiples of $\omega$ (or
$\hbar \omega$ in physical units). As we saw in Lemma~\ref{l:nondiag} the
effect of $W$ amounts to transitions between different eigenspaces of $J$.
For large $\omega$ such transitions are suppressed by the large energy
transfer. We study resolvents to make this precise, see
Lemma~\ref{l:limRes1} and Proposition~\ref{p:resolv}.

The other intuition relies on a variant of the Trotter product formula
which says that the time evolution is well approximated if one rapidly
alternates between the evolutions generated by either $H_0$ or by
$V_{\omega t}$ alone as we saw in Proposition~\ref{p:ProdFo}. A similar,
technically more convenient version are the product formulae in
Proposition~\ref{p:ProdFo2}.
In the limit $\omega \to \infty$ the latter evolution
depends only on the average $\overline{V}$ of $V_{\omega t}$. This argument
is used in the second proof of Theorem~\ref{Th1}.

\begin{proposition} \label{p:resolv}
Let $H_0$ and $V_0 = \overline{V} + W_0$ satisfy the assumptions given
in Section~\ref{s:rapRotPre} (and repeated in Theorem~\ref{Th1}) and
$J\;P_\ell = \ell\;P_\ell$. Then uniformly in $\varphi\in [0,2\pi]$
\be[limRes]
\slim_{\omega\to\infty}\;(\Ho + \omega \ell + \overline{V} + W_\varphi
-i\zeta)^{-1} =(H_0 + \overline{V} -i\zeta)^{-1}\;P_\ell\,.
\end{equation}
\end{proposition}
\begin{proof}
For $\pm\zeta > \|W_0\|=\|W_\varphi\|$ the sum in the resolvent equation
\begin{align*}
(\Ho + & \omega \ell + \overline{V} + W_\varphi -i\zeta)^{-1}\\
&=(\Ho + \omega \ell + \overline{V} -i\zeta)^{-1}\;
\sum_{n=0}^\infty \left[ -W_\varphi \;(\Ho + \omega \ell + \overline{V}
-i\zeta)^{-1}\right]^n
\end{align*}
is norm-convergent. For $\varepsilon >0$ choose $N(\varepsilon)$ such
that $\sum_{n>N(\varepsilon)} (\|W_0\|/|\zeta|)^n < \varepsilon$.
Finite products of uniformly bounded strongly convergent operators converge
as well strongly. To show the uniformity in $\varphi$ we look at the term
with $n=1$:
\begin{align*}
& W_\varphi\;(\Ho + \omega \ell + \overline{V}-i\zeta)^{-1}\;\Phi\\
&\longrightarrow \; W_\varphi\;(H_0 + \overline{V}
-i\zeta)^{-1}\;P_\ell\;\Phi\quad\text{ as }\quad \omega\to\infty.
\end{align*}
Since $W_\varphi$ is strongly continuous the set $\{W_\varphi\;\Psi\mid
\varphi \in [0,2\pi]\} $ is precompact for any given vector $\Psi$
(it can be covered by finitely
many balls of radius $\delta$ for every $\delta>0$).
We can use the strong convergence of
the next factor to the left. Similarly for higher, finite $n$.
By Lemma~\ref{l:limRes1} we get
\begin{align*}
\slim_{\omega\to\infty}\; (\Ho + & \omega \ell + \overline{V} + W_0
-i\zeta)^{-1}\\
&=(H_0 + \overline{V} -i\zeta)^{-1}\;P_\ell\;
\sum_{n=0}^\infty \left[ -W_0\;P_\ell\;(H_0 + \overline{V}
-i\zeta)^{-1}\;P_\ell\right]^n .
\end{align*}
Since $P_\ell\;W_\varphi\;P_\ell=0$ for all $\ell\in\Z$ only the term with
$n=0$ remains. This shows \eqref{e:limRes}.
\end{proof}
Now we turn to the propagator $U$ which solves the time-dependent
Schr\"odinger equation \eqref{e:Schroet} in a suitable sense, see the
discussion in Section~\ref{s:concept}. The Schr\"odinger equation and,
consequently, the propagator $U$ depend on the angular velocity $\omega$ as
a parameter.
Analogous results for classical evolutions and scattering by smooth compactly
supported potentials have been proved by Schmitz \cite{Schmitz} using
averaging methods.
\begin{theorem} \label{Th1}
Let $H_0(\cdot) \in C^1(\R^\nu, \R)$ with $H_0(\p) = h(|\p|)$ having
unbounded derivative $h'$. When the real valued multiplication operator
$V_0 = \overline{V} + W_0$ is split according to \eqref{e:Vav} we assume
that the averaged potential $\overline{V}$ satisfies for some $a<1$ and
$b<\infty$:
$\|\overline{V}\,\Psi\| \leq a\,\lVert H_0\,\Psi\rVert + b\, \lVert\Psi\rVert$
for all $\Psi$ in a domain of essential self-adjointness of $H_0$. Let
$W_0$ be bounded. Then for any $T\in\R$ (uniformly on compact intervals)
\be[limTE]
\slim_{\omega\to\infty} U(t_0+T, t_0) = e^{-iT(H_0+\overline{V})}
\end{equation}
uniformly in $t_0\in\R\,$.
\end{theorem}
The uniformity in $t_0$ is clear because $U(t_0+T, t_0) = R(\omega
t_0)\;U(T,0)\;R(\omega t_0)^*$. Since $R$ is strongly continuous and periodic
the set $\{R(\varphi)\,\Psi\mid \varphi \in\R\}$ is precompact in
$\cH$ for any vector $\Psi$. The right hand side of \eqref{e:limTE} is
rotation invariant. Therefore, it is sufficient to treat $t_0 = 0$.
{\renewcommand{\proofname}{Proof with resolvents}
\begin{proof}\hspace*{1em}\newline
We have to adjust the standard proof slightly because we do not have strong
resolvent convergence and because we need some uniformity. We take $\Phi$
from the total set of vectors with $\Phi = P_\ell\;\Phi \in \cD(H_0 +
\overline{V})$, $\ell\in\Z$, $\|\Phi\| =1$. It satisfies
$R(\omega T)\,\Phi = e^{-iT\omega\ell}\;\Phi$ and $(\Ho +\omega
\ell+\overline{V})\,\Phi = (H_0+\overline{V})\,\Phi$.

By the representation of the propagator according to the last line of
\eqref{e:propdef}
\begin{align*}
U(T;0)\;:&= e^{-iT(\Ho +\overline{V} +W_{\omega \ell})}\;R(\omega
T)\;\Phi\\
&= e^{-iT(\Ho +\omega\ell +\overline{V} +W_{\varphi})}\;\Phi
\end{align*}
for $\varphi=\omega T$. For the family of cutoff functions
$g_k(\mu):=\exp(-\mu ^2/k)$ we obtain for some $\zeta\in\R\setminus \{0\}$,
uniformly in $\omega \in\R$ and $\varphi\in[0,2\pi]$,
\begin{align*}
&\|g_k(\Ho +\omega\ell +\overline{V} +W_{\varphi})\;\Phi - \Phi\| \\
&\leq \|\,[\,g_k(\Ho +\omega\ell +\overline{V} +W_{\varphi}) -\1]\;
(\Ho +\omega\ell +\overline{V} +W_{\varphi}-i\zeta)^{-1}\|\\
&\qquad\qquad\times\;
\|(\Ho +\omega\ell +\overline{V} +W_{\varphi}-i\zeta)\;\Phi\|\\
&\leq \sup_{\mu}\left|\left(1-e^{-\mu ^2/k}\right)(\mu-i\zeta)^{-1}\right|
\;\times\; \Bigl( \|(H_0+\overline{V})\,\Phi\|+\|W_0\|+|\zeta|\Bigr).
\end{align*}
For given $\varepsilon>0$ choose $k=k(\zeta,\Phi)$ large enough such that
\begin{equation*}
\|g_k(\Ho +\omega\ell +\overline{V} +W_{\varphi})\;\Phi -\Phi\| < \varepsilon/6
\end{equation*}
and keep it fixed in the sequel. For $T$ in a compact interval $I$ the set
of functions\\
$\left\{e^{-iT\cdot}\;g_k(\cdot) \mid T\in I\right\}$ is bounded and
equicontinuous. By the Arzela-Ascoli Theorem it is precompact in the set of
bounded continuous functions tending towards zero at infinity with the
supremum norm. By the Stone-Weierstra{\ss} Theorem there are finitely many
polynomials $P_m$, $1\leq m \leq m_1$, such that
\begin{equation*}
\sup_{\mu\in\R}\left|e^{-iT\mu}\;g_k(\mu) - P_m\Bigl( (\mu-i\zeta)^{-1},
(\mu+i\zeta)^{-1}\Bigr)\right| < \varepsilon/6
\end{equation*}
for some $m=m(T)$, $T\in I$. Then
for this $m$
\begin{align} \label{e:TEres}
\Bigl\|
P_m\Bigl( \left(\Ho +\omega\ell +\overline{V}+W_{\varphi}-i\zeta\right)^{-1},
&\left(\Ho +\omega\ell+ \overline{V} +W_{\varphi}+i \zeta\right) ^{-1}
\Bigr)\;\Phi \notag\\
& \qquad\qquad\qquad
-e^{-iT(\Ho +\omega\ell +\overline{V} +W_{\varphi})}\;\Phi
\Bigr\|<\varepsilon/3
\end{align}
holds uniformly in $\omega\in\R$, $\varphi\in[0,2\pi]$, including the
special case $\omega=0$, $W=0$, i.e., functions of $(H_0 +\overline{V})$.
Finally, choose $\omega_1(\varepsilon)$ such  that for all
$\varphi\in[0,2\pi]$ and $\omega > \omega_1(\varepsilon)$
\begin{align} \label{e:limSW}
\max_{1\leq m\leq m_1} \Bigl\| P_m\Bigl( & \left(\Ho +\omega\ell +\overline{V}
+ W_{\varphi}- i\zeta\right)^{-1},\,\left(\Ho +\omega\ell +\overline{V}
+W_{\varphi}+i\zeta\right)^{-1} \Bigr)\;\Phi \notag\\
&\qquad\qquad\qquad - P_m\Bigl( \left(H_0+\overline{V}-i\zeta\right)^{-1},
\left(H_0+\overline{V}+i\zeta\right)^{-1}
\Bigr)\;\Phi\Bigr\|< \varepsilon/3
\end{align}
which is possible by Proposition~\ref{p:resolv}. Combining the estimates
\eqref{e:TEres} and \eqref{e:limSW} yields
\begin{equation*}
\Bigl\| U(T;0)\;\Phi - e^{-iT(H_0 + \overline{V})}\;\Phi\Bigr\| < \varepsilon
\end{equation*}
for all $\omega > \omega_1(\varepsilon)$ and $T\in I$.
\end{proof}
}{\renewcommand{\proofname}{Proof with the product formula}
\begin{proof}\hspace*{1em}\newline
We use the approximation of the propagator as expressed in
the product formula \eqref{e:ProdFo3} and we choose for $\varepsilon >0$
some large fixed $n$ with $n> (T\,\|W_0\|)^2 \,e^{(T\,\|W_0\|)}
\,/\varepsilon$. Then
\begin{align*}
&\left\|\left( U(t_0+T;t_0) - e^{-iT(H_0 +
\overline{V})}\right)\;\Phi\right\|\\
&\quad
\leq \frac\varepsilon 2 + \left\|\left(\prod_{k=0}^{n-1}\left[ e^{-iT(H_0
+\overline{V})/n}\;
\;\ttu\!\left(\frac{(k+1)T}{n}, \frac{kT}{n}\right) \right]
- e^{-iT(H_0 +
\overline{V})}\right)\;\Phi\right\|\\
&\quad
\leq \frac\varepsilon 2 +
\sum_{k=0}^{n-1}\left\|\left\{\ttu\!\left(\frac{(k+1)T}{n},
\frac{kT}{n}\right)-\1 \right\}\;e^{-ikT(H_0 +
\overline{V})/n}\;\Phi\right\|\\
&\quad
\leq \frac\varepsilon 2 + \sum_{k=0}^{n-1}\left\|\left\{
\int_{0}^{T/n} ds\;e^{is(H_0+\overline{V})}
\;W_{\omega( s+kT/n)}\;e^{-is(H_0+\overline{V})}
\right\}\;e^{-ikT(H_0 +
\overline{V})/n}\;\Phi\right\|
\end{align*}
Now we fix $\Phi$ from the total set of vectors with
$\Phi = P_\ell\;\Phi$ for
some $\ell\in\Z$. Note that due to strong continuity of $e^{-i\tau(H_0 +
\overline{V})}$ the set of vectors $\{e^{-i\tau(H_0
+\overline{V})}\;\Phi\mid \tau\in I\}$ is precompact for any compact
interval $I$. The same is true when the bounded operator $W_0$ is
applied to this set.

Due to rotational invariance of $(H_0+\overline{V})$ the projector
$P_\ell$ can be moved to the right of $W$ and we obtain for a summand
in the last formula
\begin{equation*}
\left\|\left\{ \int_{0}^{T/n} ds\;e^{is(H_0+\overline{V})} \; e^{-i\omega (J-\ell)( s+kT/n)}\;
W_0\;P_\ell\;e^{-is(H_0+\overline{V})} \right\}\;e^{-ikT(H_0 + \overline{V})/n}\;\Phi\right\|
\end{equation*}
By equation \eqref{e:nondiag} $W_0\;P_\ell=\sum_{j\in\Z,\,j\neq \ell}
P_j\;W_0\;P_\ell$ and the precompactness implies that only finitely many $j's$ matter. For all
$\tau\in I$
\begin{equation*}
\biggl\| W_0\;P_\ell\;e^{-i\tau (H_0+\overline{V})}\;\Phi - \sum_{j\neq \ell}^{\text{finite}}
P_j\;W_0\;P_\ell\;e^{-i\tau (H_0+\overline{V})}\;\Phi\biggr\|< \varepsilon/4n.
\end{equation*}
It remains to estimate a finite sum of terms with $j\neq\ell$
\begin{equation*}
\biggl\| \int_{0}^{T/n} ds\;e^{-i\omega (j-\ell)s}\; e^{is(H_0+\overline{V})}\;
P_j\;W_0\;P_\ell\;e^{-is(H_0+\overline{V})}\;e^{-ikT(H_0 + \overline{V})/n}\;\Phi\biggr\|.
\end{equation*}
The integrands are bounded continuous vector valued functions of $s$ and, consequently, are
integrable when restricted to the interval $[0,T/n]$. By the Riemann-Lebesgue Lemma their Fourier
transform tends to zero as $\omega\to\infty$. There is $\omega_1(\varepsilon)$ such that the sum
is bounded by $\varepsilon/4$ for $\omega>\omega_1(\varepsilon)$. This shows that
\begin{equation*}
\left\|\left( U(t_0+T;t_0) - e^{-iT(H_0 +
\overline{V})}\right)\;\Phi\right\|<\varepsilon\qquad\text{for}\quad \omega>\omega_1(\varepsilon).
\end{equation*}
This concludes the second proof of \eqref{e:limTE}.
\end{proof}
}

\section{The self-adjoint sum $\Ho + V_0$} \label{s:sum}

For the special case $\omega =0$ the self-adjoint operator or form
sum $H_0 + V_0$ has
been studied extensively, mainly by methods of perturbation theory,
see, e.g., \cite{RS2}. Here we consider only the case $\omega \ne 0$
(unless otherwise stated) for $\Ho$ as given in
equations~\eqref{e:Hom1} and \eqref{e:Hom}.

Following Tip \cite{Tip} we derived in \cite[Lemma~3.1]{Goa}
that $V_0$ is bounded relative
to $\Ho$ with bound less than one if $(1+|\x|^2)\, V_0$ is bounded
relative to $H_0=|\p|^2/2m$ with bound less than one. The decay is important
only for singular potentials, an arbitrary bounded part can always be
added. In this section we treat as an example the special case of
dimension $\nu=2$ and $H_0(\p) =|\p|^2/2$ (mass $m=1$ in adjusted
units). We will show that even for
locally square integrable potentials no decay towards infinity is
needed. Higher dimensions and more general free Hamiltonians will be
addressed in a forthcoming paper.

While the global properties of $H_0$ and $\Ho$ differ very much it is
easier to control their difference locally. Therefore, we begin with
potentials of compact support.

In two dimensions let $\x^\perp = (-x_2, x_1)^{\text{tr}}$. Then
$J=\x\wedge\p = \x^\perp \cdot \p$.
\begin{lemma}
Let $V\in L^2(\R^2)$ have compact support in the unit square centered
at $\ox \in \Z^2$ and let $\chi \in C_0^\infty(\R^2)$
satisfy $\chi(\x-\ox)=1$ in a neighborhood of the support of $V$.
Then for any $a>0$ there is a $b=b(a)<\infty$ such that for $\Psi$
with $\psi(\x) \in C_0^\infty(\R^2)$
\begin{align}
\| V \;\chi (\cdot -\ox) \; \Psi\| & \leq a\,\|(H_0 -\omega \oxp \p )\;\chi(\cdot
-\ox)\;\Psi\| +b \;\|\chi(\cdot -\ox)\;\Psi\|,\label{e:VergLin} \\[1ex]
\| V\;\chi(\cdot -\ox)\;\Psi\| & \leq a\,\|(H_0 -\omega J)\;\chi(\cdot-\ox)\;\Psi\|
+b\;\|\chi(\cdot -\ox)\;\Psi\|.\label{e:VergJ}
\end{align}
The bounds $a$ and $b$ depend on $\|V\|_2$, but they can be chosen independent of $\ox$.
\end{lemma}
For fixed $\ox$ equation~\eqref{e:VergLin} is well known in any
dimension. The uniformity in  $\ox$ is important here.
\begin{proof}
With $\op := \omega\oxp$ we have $H_0(\p) -\omega\oxp \p=
H_0(\p -\op) -|\op|^2/2$. For any $a>0$ we estimate the
$L^2$-norm of the following function of $\p$:
\begin{align*}
\Bigl\|\bigl(a[(\p-\op)^2 -|\op|^2/2\,] -i/a\bigr)^{-1}\Bigr\|_2^2
&=\int \frac{d\p}{\{a\,[\,(\p-\op)^2 -|\op|^2/2\,]\,\}^2 +
1/a^2} \\[1ex]
&=\pi\int_{|\op|^2/2}^\infty \frac{du}{u^2+1} \\
&=\pi\{\pi/2 + \arctan(|\op|^2/2)\}\leq \pi^2
\end{align*}
where we have used polar coordinates around $\op$ and
$u=a^2\,[\,(\p-\op)^2 -\lambda\,]$. This gives the uniformity in
$\op$, the remaining proof is standard. Denoting by
$(\widehat{\chi\psi})(\p)$ the Fourier transform of
$(\chi\psi)(\x):=\chi(\x-\ox)\,\psi(\x)$ we estimate
\begin{align*}
& \|(\chi\psi)\|_\infty \leq (1/2\pi)\;\|(\widehat{\chi\psi})\|_1 \\
&\leq \frac{1}{2\pi}\;\left\| \frac{1}{a[(\p-\op)^2 -|\op|^2/2\,] -i/a}
\right\|_2 \;\left\| (a[(\p-\op)^2 -|\op|^2/2\,]
-i/a)\;(\widehat{\chi\psi})\right\|_2 \\[1ex]
&\leq a \|(H_0 -\op\p)\;(\chi\psi)\|_2 + (1/2a)\; \|(\chi\psi)\|_2\, .
\end{align*}
With $\|V\;(\chi\psi)\|_2 \leq \|V\|_2\;\|(\chi\psi)\|_\infty$ this
shows the estimate \eqref{e:VergLin} uniformly in $\ox$. Then
\eqref{e:VergJ} follows easily from the observation that
\begin{equation*}
(J-\oxp\p)\;\chi(\cdot-\ox) = (\x-\ox)\cdot \nabla\chi(\cdot-\ox) +
(\x-\ox)\,\chi(\cdot-\ox)\cdot\p
\end{equation*}
with uniformly bounded functions of $\x$.
\end{proof}
Now we split a potential $V\in L^2_{\text{loc}}$ into four parts. The first of them, $V^{(1)}$,
has its support only in those unit squares which are centered at those $\ox\in\Z^2$ which have
\textit{even} integers as coordinates. The remaining three parts have both coordinates of the
centers odd or one even and the other odd. In each of the four components each unit square which
belongs to the support is well separated from all others.
Now we choose a decomposition of the identity
\begin{equation*}
\sum_{\ox\in (2\Z)^2} \;[\,\chi(\cdot-\ox)\,]^2 =1
\end{equation*}
where $\chi \in C_0^\infty(\R^2)$ and $\chi(\x)=1$ in a neighborhood of the unit square around the
origin. This decomposition splits the potential $V^{(1)}$ into pieces which coincide with $V$ in
one unit square and are zero outside of it. For the other components of the potential we use
decompositions which are shifted by $(0,1)$, $(1,0)$, or $(1,1)$, respectively.

For $V\in L^2_{\text{loc,\,unif}}$ the $L^2$-norms of the restrictions to arbitrary unit squares
are uniformly bounded. This applies, in particular, to all parts of $V$ constructed above.
\begin{theorem} \label{Th2}
Any $V\in L^2_{\text{\rm loc,\,unif}}(\R^2)$ is bounded relative to $\Ho$ with relative bound
zero. In particular, $(\Ho + V)$ is essentially self-adjoint on any core of $\Ho$.
\end{theorem}
\begin{proof}
Morgan has shown in \cite[Theorem~2.3]{Morgan} that \eqref{e:VergJ}
implies
\begin{equation*}
\|V^{(1)}\;\Psi\| = \|V^{(1)}\;\sum_{\ox\in (2\Z)^2}\chi(\cdot-\ox)\;\Psi\| \leq a\,\|\Ho \Psi\| +
b\,\|\Psi\|
\end{equation*}
and analogously for the other three components.
\end{proof}
\bibliographystyle{amsalpha}

\end{document}